\documentclass{article}
\usepackage{amsbsy}
\usepackage{icmpprep,epsf}
\usepackage{wrapfig}

\newenvironment{lecture}[3]{
\clearpage
\begin{center}
{\Large\bf #1\addcontentsline{toc}{section}{#1 (#2)}}\\ [1em]
{\large\bf #2}\\ [1ex]
{\large\em #3}
\end{center}
}{
\setcounter{section}{0}
\setcounter{subsection}{0}
\setcounter{equation}{0}
\setcounter{figure}{0}
\setcounter{table}{0}
\setcounter{footnote}{0}
}
\begin{document}
\begin{lecture}
{
%
%
     THERMODYNAMICS OF A PSEUDOSPIN-ELECTRON MODEL
}
{
%
%
  Tabunshchyk K.V.
}
%
%
{Institute for Condensed Matter Physics Nat. Acad. Sci. Ukr.
            1 Svientsitskii Str., UA--79011 Lviv, Ukraine}
%

\newcommand{\bk}{\boldsymbol {k}}
\newcommand{\bq}{\boldsymbol {q}}
\newcommand{\fl}{\hspace*{-1pc}}
\newcommand{\fll}{\hspace*{-3pc}}
\newcommand{\rs}{\rm\scriptscriptstyle}

\section*{Abstract}
{\small
 The purpose of this article is to present the
thermodynamics of the pseudospin electron (PE) model in the case
of the different type interactions between pseudospins.
 First, we provide an overview of the results of works which deal with the
theoretical investigation of the PE model with the inclusion of
the direct pseudospin-pseudospin interaction (but without the
electron transfer).
 Second, we present the results of the investigation of the model in
the case of the absence of the direct pseudospin-pseudospin
interaction and Hubbard correlation, when interaction between
pseudospins via conducting electron is done.}

\newpage

\section*{Introduction}

 Pseudospin-electron (PE) model is one of theoretical models which
considers the interaction of electrons with local lattice vibrations
where an anharmonic variables are represented by pseudospins.
 The theoretical investigation of the PE model is an enduring subject
of interest at the quantum statistics department.

 The model is used to describe the strongly correlated electrons
of CuO$_2$ sheets coupled with the vibrational states of apex oxygen
ions O$_{\rm IV}$ (which move in the double-well potential)
in YBaCuO type high-T$_{\rm c}$ superconductors (HTSC) \cite{1}.
 Recently a similar model has been applied for investigation of the
proton-electron interaction in molecular and crystalline systems with
hydrogen bonds \cite{2}.

 The model Hamiltonian is the following:
\begin{eqnarray}
 \label{1}
 &&H=\sum\limits_iH_i +\sum\limits_{ij\sigma}t_{ij} a_{i\sigma}^+
  a_{i\sigma}-\frac 12 \sum\limits_{ij}J_{ij}S_i^zS_j^z,\\
 &&H_i=Un_{i\uparrow}n_{i\downarrow}-\mu\sum\limits_{\sigma}n_{i\sigma }
  +g\sum\limits_{\sigma}n_{i\sigma}S_i^z-hS_i^z,
 \nonumber
\end{eqnarray}
where the strong single-site electron correlation $U$,
interaction with the anharmonic mode ($g$-term), and the energy
of the anharmonic potential asymmetry ($h$-term) are included
in the single-site part.
 Hamiltonian~(\ref{1}) also contains terms, which describe electron
transfer $t_{ij}$ and direct interaction between pseudospins $J_{ij}$.
 The energy of the electron states at the lattice site is accounted from
the level of chemical potential $\mu$.

 In the case of strong coupling ($g\gg t$) and correlation ($U\gg t$) the
perturbation theory can not be applied.
 It is reasonable to include these one in zero order Hamiltonian
 (single-site Hamiltonian $H_i$).
 Its eigenfunctions are build of the vectors
$\left| n_{i\uparrow},n_{i\downarrow},S_i^z\right\rangle$, which
form the full unit cell state basis \cite{3}:
\begin{eqnarray}
\label{2}
\fll
&&|1\rangle =|0,0,\frac 12\rangle,\quad
  |2\rangle =|1,1,\frac 12\rangle,\quad
  |3\rangle =|0,1,\frac 12\rangle,\quad
  |4\rangle =|1,0,\frac 12\rangle,\\
\fll
&&|\tilde{1}\rangle =|0,0,-\frac 12\rangle,\,
  |\tilde{2}\rangle =|1,1,-\frac 12\rangle,\,
  |\tilde{3}\rangle =|0,1,-\frac 12\rangle,\;
  |\tilde{4}\rangle =|1,0,-\frac 12\rangle.
  \nonumber
\end{eqnarray}

 In the early works the main attention at the investigation of
this model has been paid to the examination of electron states,
effective el\-ec\-tron-el\-ec\-tron interaction, to the elucidation of
additional possibilities of occurrence of superconducting pair correlations.
 On the basis of PE model a possible connection between the
superconductivity and lattice instability of the ferroelectric
type in HTSC has been discussed \cite{Tang,Frick}.
 A series of works has been carried out in which the pseudospin
$\left\langle S^zS^z\right\rangle$, mixed $\left\langle S^zn\right\rangle$
and charge  $\left\langle nn\right\rangle $ correlation functions
were calculated.
 It has been shown with the use of the generalized random phase
approximation (GRPA) in the limit $U\to\infty$ \cite{3,4}, that there
exists a possibility of divergences of these functions at some
values of temperature.
 This effect was interpreted as a manifestation of dielectric instability or
ferroelectric type anomaly.
 The tendency to the spatially modulated charge and pseudospin ordering
at the certain model parameter values was found out.
 The analysis of ferroelectric type instabilities in the two-sublattice model
of high temperature superconducting systems has been made
\cite{Danyliv}.
 The influence of oxygen nonstoichiometry on localization of apex
oxygen in YBa$_2$Cu$_3$O$_{7-x}$ type crystals is studied in the
work \cite{Velychko}.

 The purpose of this article (lecture) is to present the thermodynamics of
the PE model in the case of the different type interactions
between pseudospins.
 First, we provide an overview of the results of works which deal with the
theoretical investigation of the PE model with the
inclusion of the {\em direct pseudospin-pseudospin interaction}
(but without the electron transfer ($t_{ij}=0$)).
 Second, we present the results of the investigation of the model in
the case of the absence of the direct pseudospin-pseudospin interaction
and Hubbard correlation ($J_{ij}=0$, $U=0$), when {\em interaction
between pseudospins via conducting electron is done}.

\section{Direct interaction between pseudospins.}

\subsection{Ferroelectric type interaction.}

 The work \cite{5} is devoted to the study of the PE model in the case of
zero electron transfer $(t_{ij}=0)$.
 The direct interaction between pseudospins is taken into account.
 It is supposed to be a long-ranged 
that allows one to use the mean field approximation (MFA).
 In this approximation, the model Hamiltonian has the following form:
$$
H{=}\sum\limits_i\tilde{H}_i{+}\frac N2 J\eta^2,\;\,
\tilde{H}_i{=}{-}\mu\sum\limits_{\sigma}n_{i\sigma}{+}Un_{i\uparrow}
 n_{i\downarrow}{+}g\sum\limits_{\sigma}n_{i\sigma}S_i^z{-}(h{+}J\eta)S_i^z.
$$

 The interaction $(J_{ij}\sim~J/N)$ is taken as the ferroelectric type one;
the order parameter $\eta=\langle S_{i}^{z}\rangle$ does not depend
on the unit cell index.

 Grand canonical potential and partition function of the model,
calculated per one lattice site are equal to
\begin{eqnarray}
  \label{3}
 &&\fl
  \frac{\Omega}{N}=-T\ln Z_i+\frac 12 J\eta^2,\\
 &&\fl
  Z_i=2\left[\cosh\beta h_n{+}{\rm e}^{-\beta(U-2\mu)}\cosh\beta
  (h_n{-}g){+}2{\rm e}^{\beta\mu}\cosh\beta\left(h_n{-}
  \frac g2\right)\right],
  \nonumber
\end{eqnarray}
where $h_n=h/2+J\eta/2$.

 Then, all thermodynamic properties can be derived from the presented
formulae (\ref{3}).

 The average number of electrons is determined as follows \cite{5}:
\begin{equation}
  \label{4}
 -\frac 1N\left(\frac{\partial\Omega}{\partial\mu}\right)_{T,h,\mu}
 =\left\langle\frac 1N\sum\limits_in_i\right\rangle
 \equiv \langle n\rangle.
\end{equation}

 The equation for the order parameter is obtained from the thermodynamical
relation \cite{5}:
\begin{equation}
 \label{5}
 \left(\frac{\partial\Omega}{\partial\eta}\right)_{T,h,\mu}=0.
\end{equation}
 For the investigation of equilibrium conditions one should
separate two regimes: $\mu=const$ and $n=const$.
 We would like to note that hereafter we shall demonstrate the results of
numerical investigation which show the main features of the considered model.

\subsubsection{$\mu=const$ regime.}

 The $\mu=const$ regime corresponds to the case when, for example, charge
redistribution between the conducting sheets CuO$_2$ and other structural
elements is allowed.
 For this regime the equilibrium is defined by the minimum of the grand
canonical potential that form an equation for pseudospin mean
value (\ref{5}) and expression for $\langle n\rangle$ (\ref{4}).

 At some regions of $h$ values the field dependencies $\eta (h)$
possesses $S$~-~like behaviour Fig.~\ref{fig1} (the first order phase
transition with the jump of order parameter and
electron concentration take place at the change of field $h$).
 The phase transition point is presented by a crossing point
on the dependence $\Omega (h)$.
\begin{figure}[hbt]
 \centerline{
  \epsfysize 3.cm\epsfbox{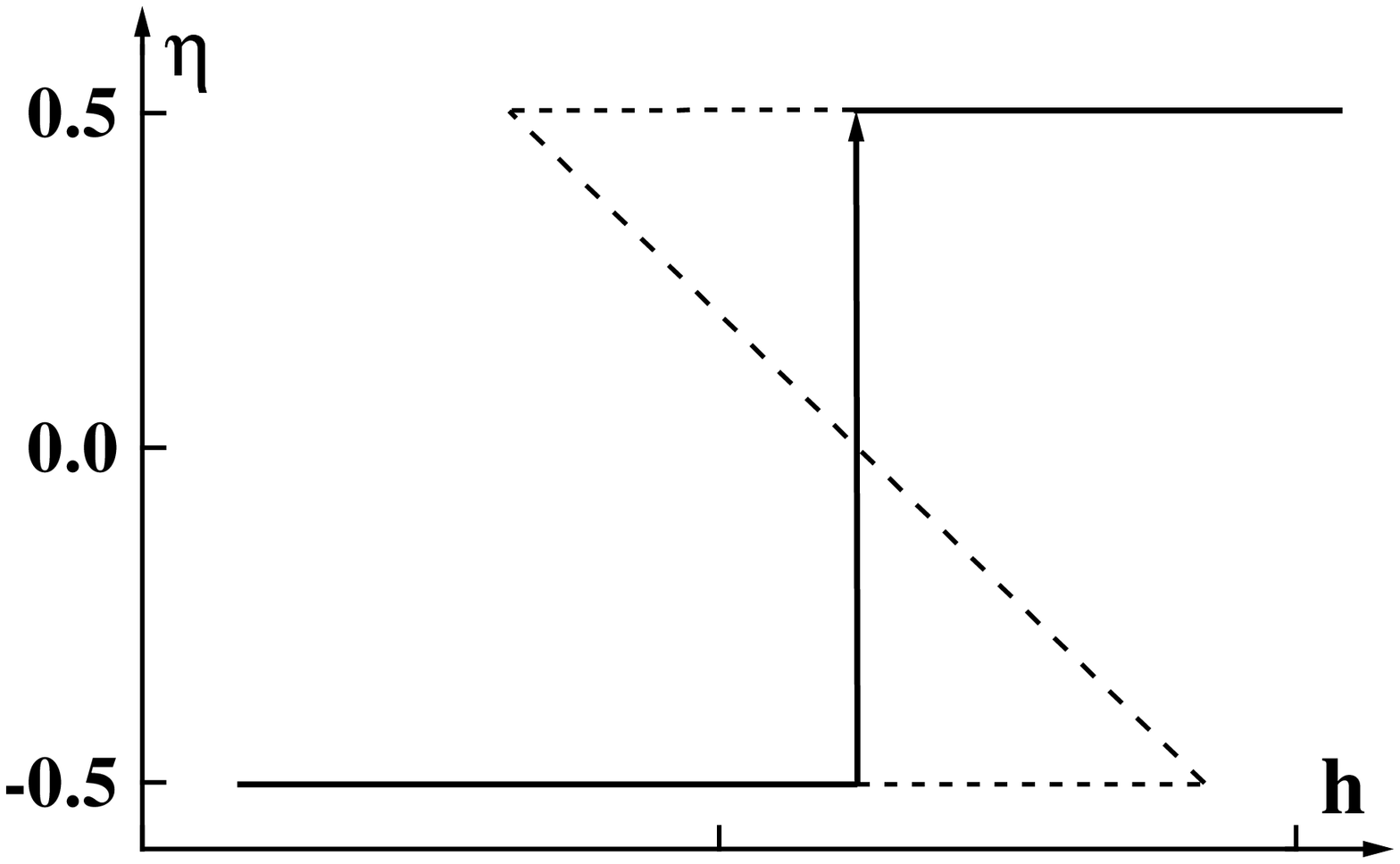}
  \qquad \epsfysize 3.cm\epsfbox{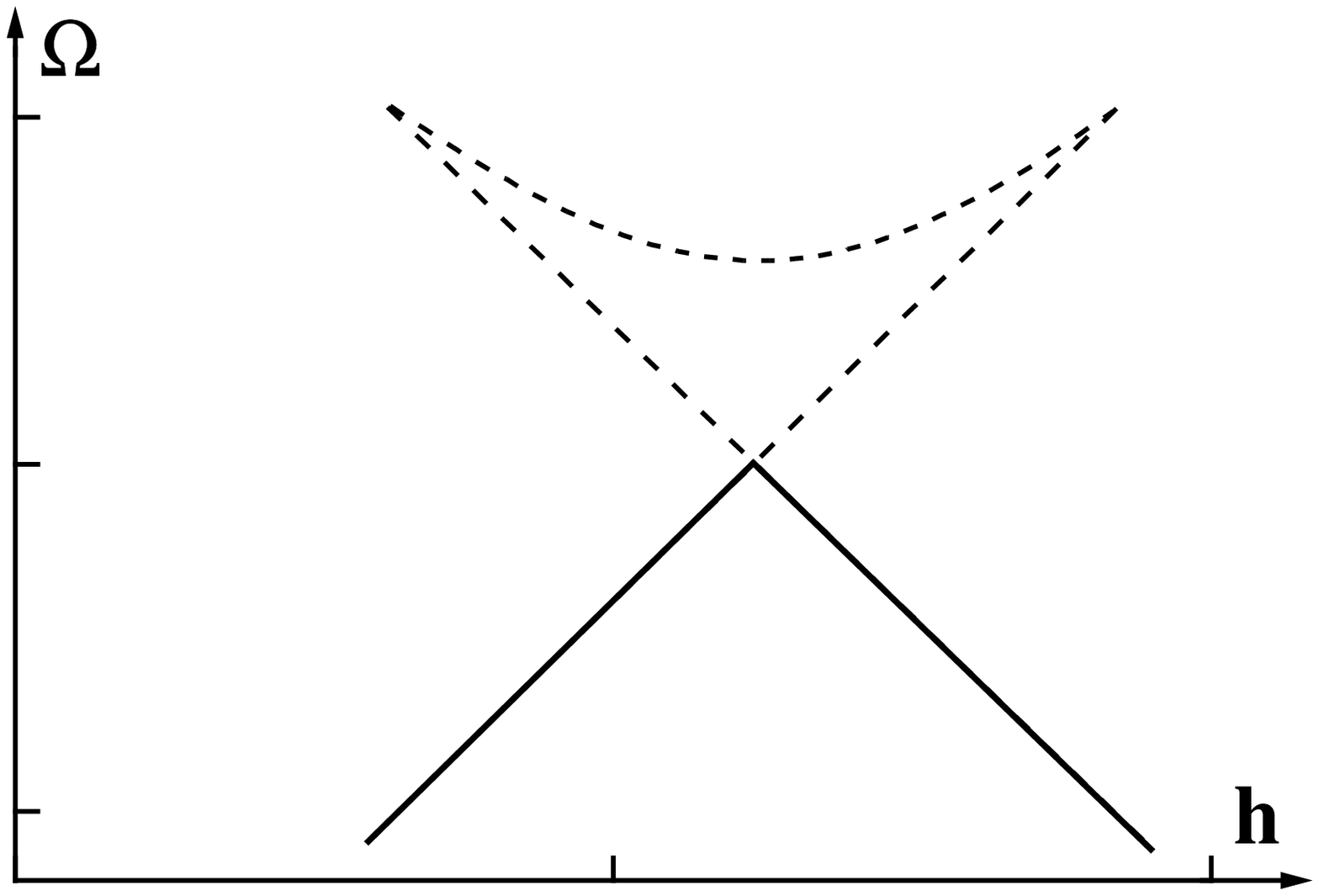}
  }
  \caption{Field dependencies of the order parameter
           and grand canonical potential ($T=0$).}
 \label{fig1}
\end{figure}

 The phase diagram $T_c-h$ is shown in Fig.~\ref{fig2}.
 One can see that with respect to Ising model the phase coexistence
curve is shifted in field and distorted from the vertical line and
hence the possibility of the first order phase transition with the
temperature change exists in PE model.
\begin{figure}[htbp]
 \centerline{\quad
  \epsfysize 3.cm\epsfbox{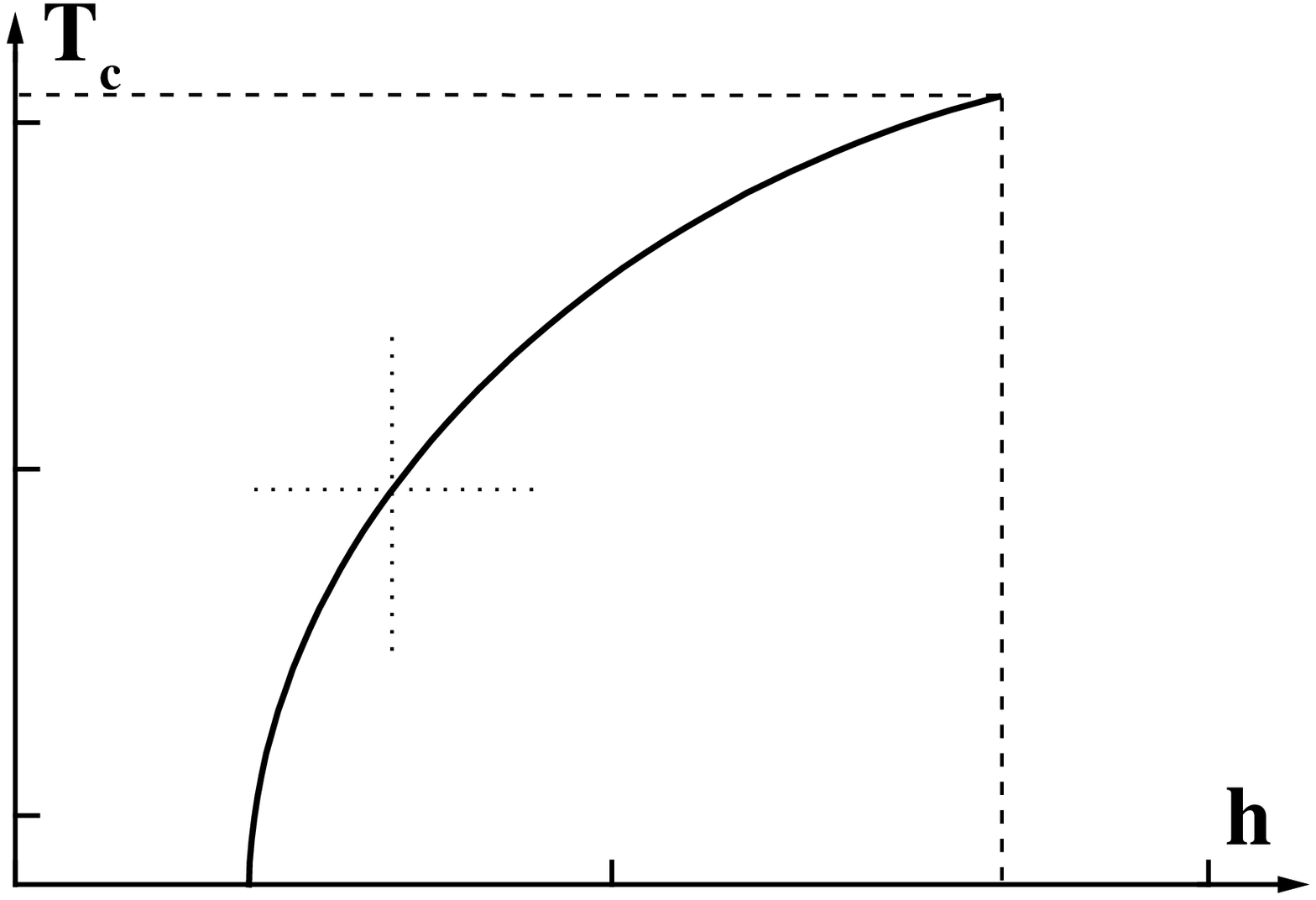}
  \qquad  \raisebox{-.2cm}{
  \epsfysize 3.2cm\epsfbox{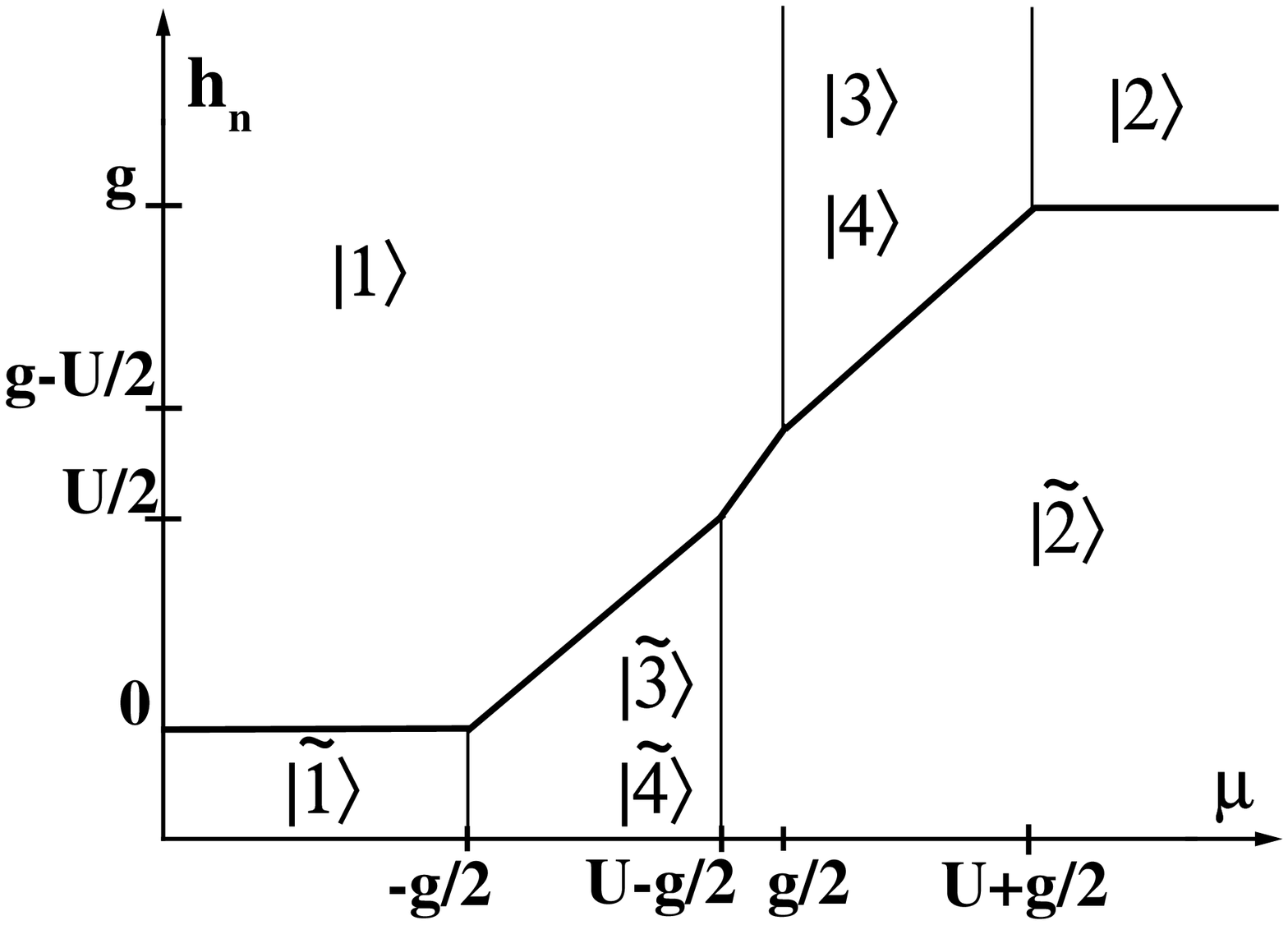}
  }
  }
  \caption{$T_c-h$ and $\mu-h$ phase diagrams.}
 \label{fig2}
\end{figure}

 The phase diagram $\mu-h$ (Fig.~\ref{fig2}) indicate stability
regions for states with $\eta=\pm 1/2$ ($U<g$ and $T=0$).
 The form of diagram depends on the relation between $U$ and $g$ parameter
values \cite{5}.
 Transitions between regions
$|r\rangle\leftrightarrow |p\rangle$,
$|\tilde r\rangle\leftrightarrow |\tilde p\rangle$
lead to the change of the average number of electrons only.
 At transitions
$|r\rangle\leftrightarrow |\tilde r\rangle$
the flipping of pseudospin takes place, and at
$|r\rangle\leftrightarrow |\tilde p\rangle$ $(r\neq p)$
both processes occur.

 Hence, the possibility of the first order phase transition with the
change of field $h$ and/or chemical potential $\mu$ takes place and
is shown by thick line on phase diagram in Fig.~\ref{fig2}.

\subsubsection{$n=const$ regime.}
\begin{wrapfigure}[9]{r}{4.2cm}
 \label{fig3}
 \raisebox{.6cm}[2.cm][.cm]{\epsfxsize 4.2cm\epsfbox{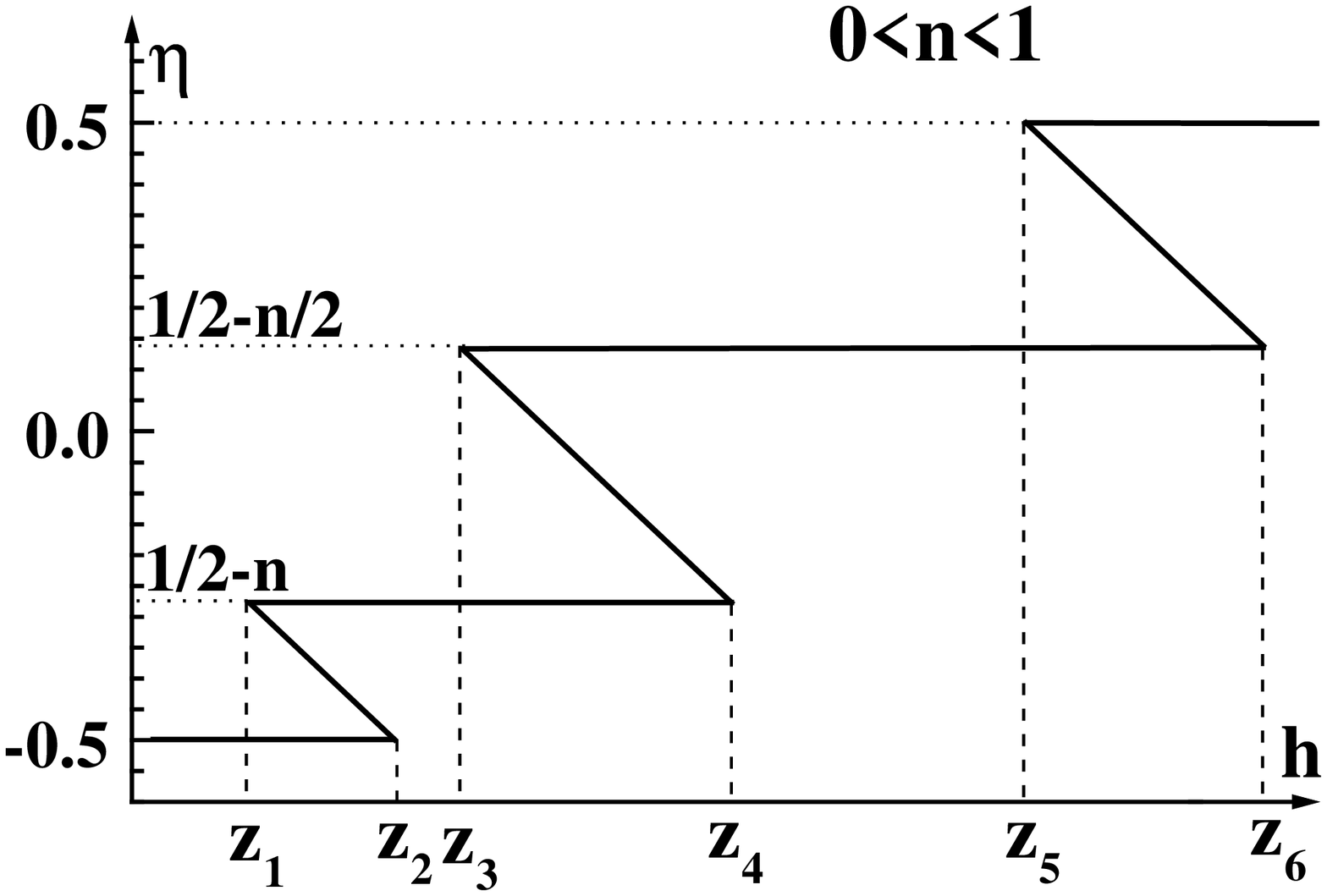}}
 \parbox[c]{4.cm}{\vspace*{-.3cm}
 Figure 3. Field depen\-den\-cy of the order pa\-ra\-meter ($T=~0$).}
\end{wrapfigure}
 In the regime of the fixed electron concentration value the equilibrium is
determined by the minimum of free energy $F=\Omega+\mu N$ and form
a set of equations (\ref{4}), (\ref{5}) for the chemical potential
and order parameter.

 Typical example of the $\eta (h)$ dependence is shown in Fig.~3
which corresponds to the $\langle n\rangle$ value in the interval
\mbox{$0\le n\le 1$}.
 Phases \mbox{$\eta=-\frac 12$} (phase~1),
\mbox{$\eta=\frac 12-n$} (phase~2),
\mbox{$\eta=\frac 12-\frac n2$} (pha\-se~3),
\mbox{$\eta=\frac 12$} (phase~4) exist
between phase transition points (which is determined according to the
 Maxwell rule from the range of $S$-like behaviour (between the spinodal
points~$Z_i$))
and outside of them.
 At the change of the model parameter values the regions, where metastable
phases exist, can overlap, some phase transitions disappear (some
intermediate phases can not be realized).
 In case $1\le n\le 2$ the dependence $\eta (h)$ is generally similar.
 The phase 3 and phase 2' at $\eta=\frac{3}{2}-n$, which now appears instead
of phase 2, may play the role of the intermediate phases.

\begin{wrapfigure}[10]{r}{4.2cm}
 \label{fig4}
 \raisebox{-.6cm}[2.cm][.7cm]{\epsfxsize 4.2cm\epsfbox{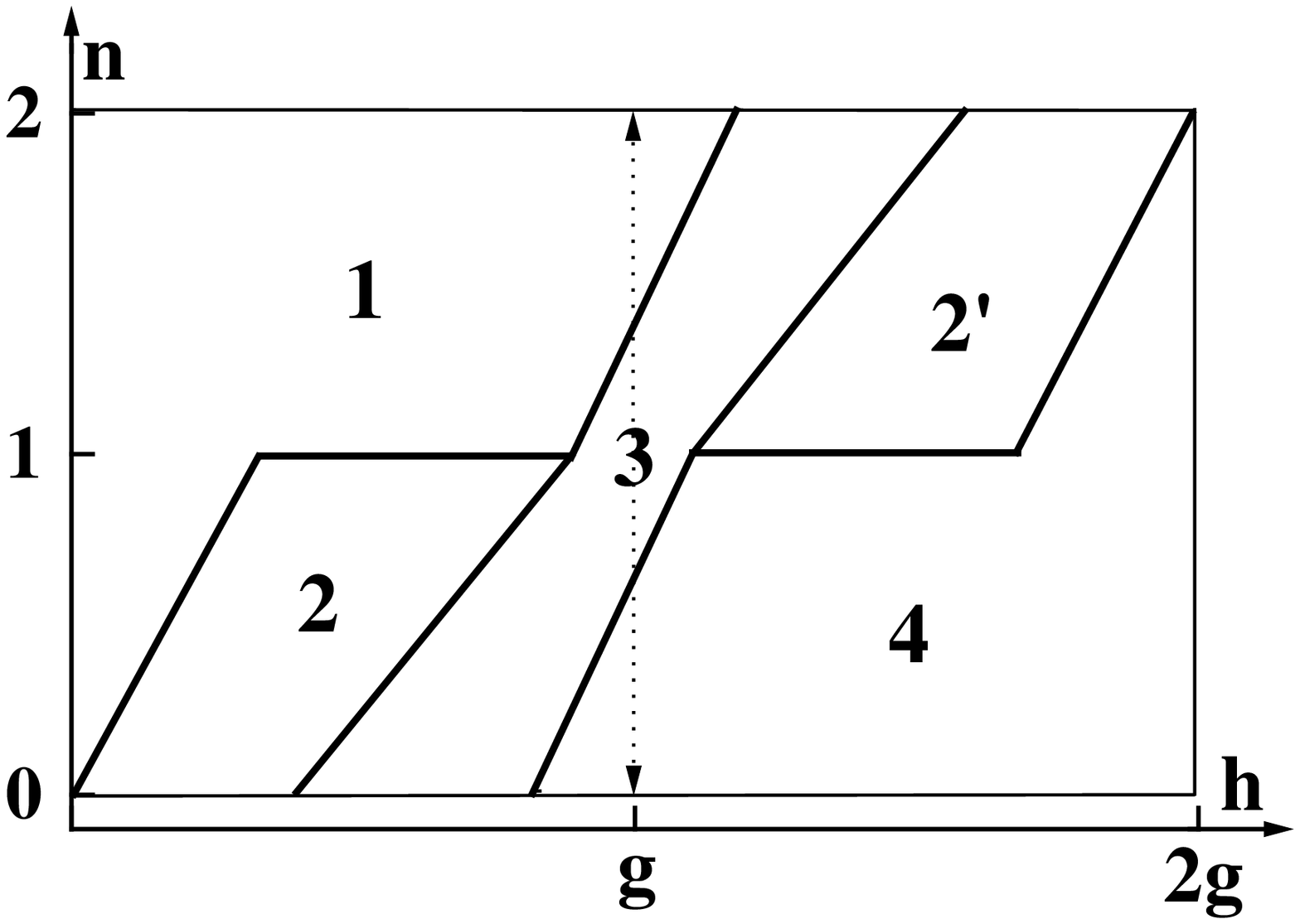}}
 \parbox[c]{4.1cm}{Figure 4. $n-h$ phase transition diagram ($T=0$).}
\end{wrapfigure}
 On the phase diagram $n-h$ Fig.~4 the thick solid line indicates the phase
coexistence curve and hence the possibility of the first order phase
transition with the change of the longitudinal field $h$ and/or electron
concentration $\langle n\rangle$ takes place.

 More detail analyse of a free energy behaviour shows that the above
presented (in this paragraph) results are not realized.
 The investigation of the equilibrium conditions shows that
the first order phase transition transforms into the phase separation.
 One can see regions where state with homogenous distribution of particles
is unstable ($d\mu/dn\le 0$), and the phase separation into the
regions with different concentrations exists (Fig.~5).
 The phase diagram $n-h$ (Fig.~5) illustrates the separation phases.
\begin{figure}[htbp]
 \centerline{
  \epsfysize 3.cm\epsfbox{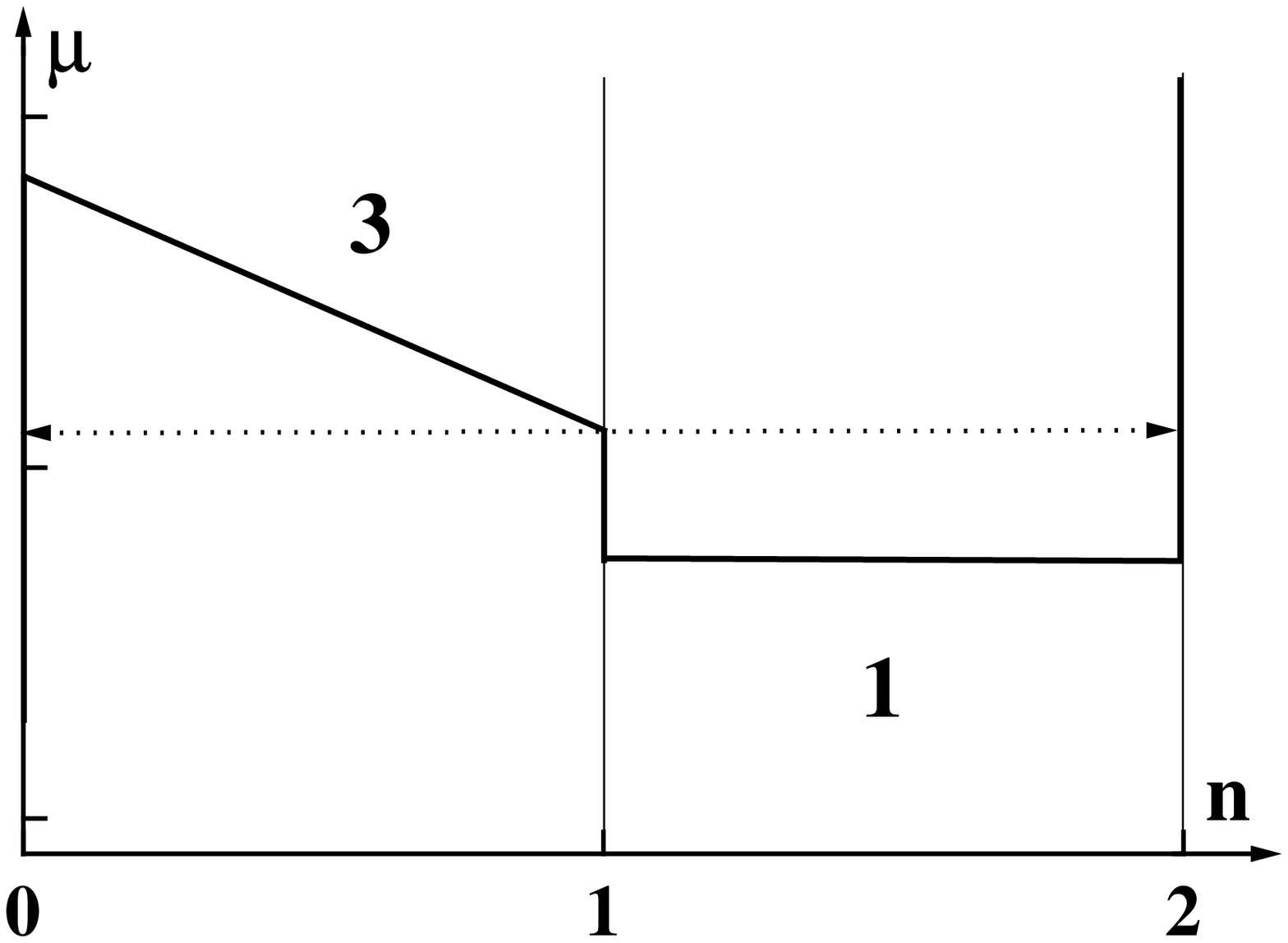}
  \qquad  \epsfysize 3.cm\epsfbox{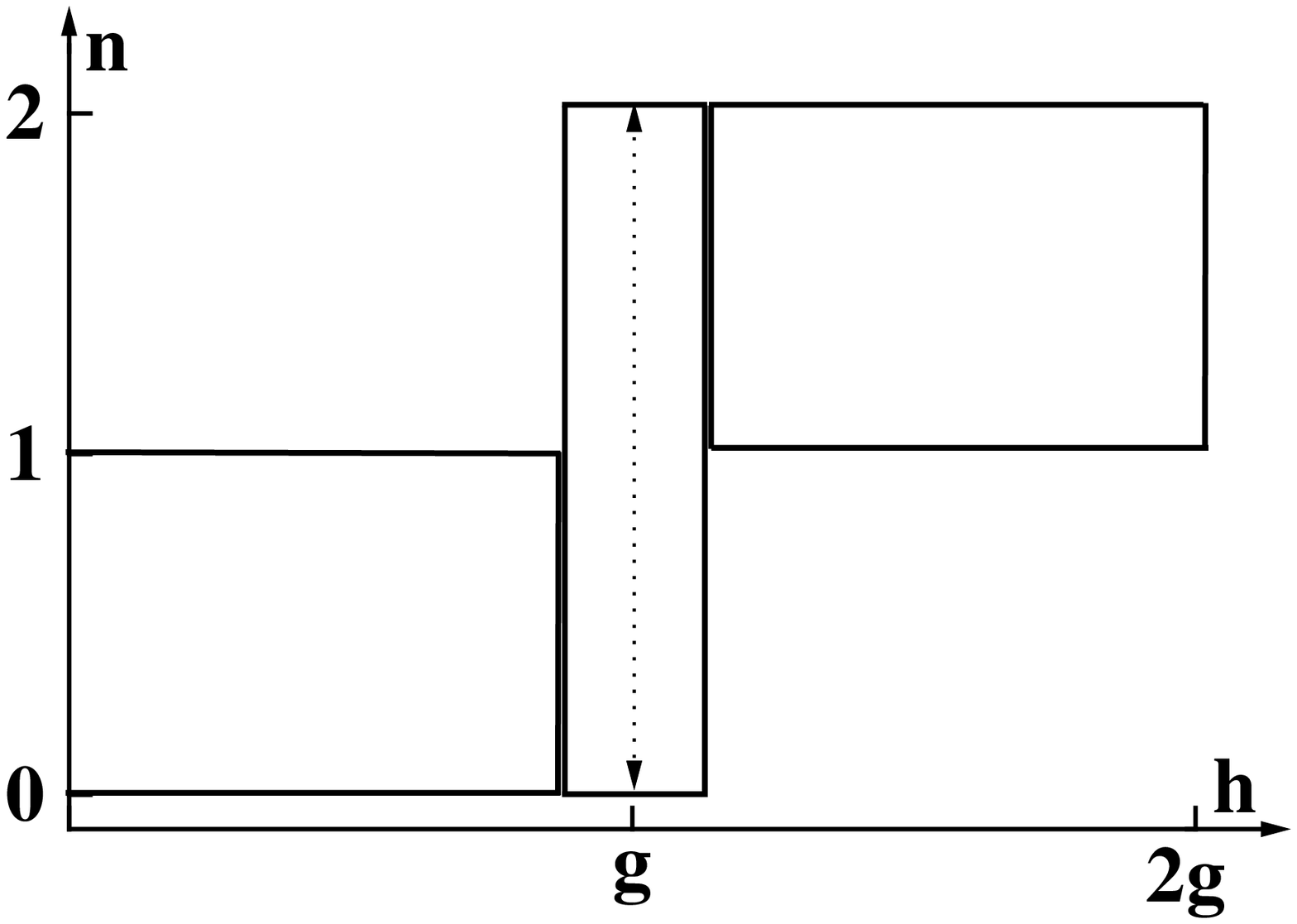}
  }
  Figure 5. Dependence of the chemical potential $\mu$ on the electron
           concentration and $n-h$ phase separation diagram ($T=0$).
 \label{fig5}
\end{figure}

 The phase 3 splits into phase 4 (with concentration $n=0$, order parameter
$\eta=\frac{1}{2}$) and phase 1 ($n=2$, $\eta=-\frac{1}{2}$) with
weight coefficients $1-n$ and $n$ respectively (thin dotted lines
in Fig.~4 and Fig.~5).

 Therefore, more convenient and thermodynamically stable is the
pha\-se separated state, which is the mixture of states with different
electron concentrations and different values of order parameters.

\subsection{Antiferroelectric type interaction.}

In the case of antiferroelectric type interaction it is convenient to
introduce two kinds of sites (A-sites, B-sites). These corresponds to the
doubling of the lattice period \cite{6}.

Within the framework of the MFA we shall write:

\begin{equation}
\label{6}
 S_A^zS_B^z=-\eta_A\eta_B+\eta_AS_B^z+\eta_BS_A^z,
\end{equation}
where $\eta_A=\langle S_{A}^z\rangle$, $\eta_B=\langle S_{B}^z\rangle$.

Then, we obtain the following expression for the model Hamiltonian:
\begin{eqnarray}
\label{7}
H&=&\sum_i\tilde H_{iA}+\sum_j\tilde H_{jB}+
    N\left\{\frac {J_1}2\eta_A\eta_B+\frac {J_2}4(\eta_A^2+\eta_B^2)\right\},
    \nonumber\\
\tilde H_{iA}&=&-\mu(n_{iA\downarrow}+n_{iA\uparrow})+
    Un_{iA\downarrow}n_{iA\uparrow}
    +g(n_{iA\downarrow}+n_{iA\uparrow})S_{iA}^z-\nonumber\\
&&-(h+J_1\eta_B+J_2\eta_A)S_{iA}^z, \nonumber\\
\tilde H_{jB}&=&\tilde H_{jA}\big|_{A\leftrightarrow B}.
\end{eqnarray}
 The Hilbert space forms as a direct product of the eigenfunctions (\ref{2}) for
$\tilde H_{A}$ and $\tilde H_{B}$ operators (\ref{7})
$\left\{|n_{iA\uparrow} N_{iA\downarrow}, S_{iA}^z\rangle\right\}
\oplus \left\{|n_{iB\uparrow} N_{iB\downarrow},
S_{iB}^z\rangle\right\}$.
 The analytical consideration in this case in general is very similar to the
previous (ferromagnetic interaction) one, but formulae are more complicated
(see in details \cite{6}).

 Grand canonical potential can be written in the form:
\begin{eqnarray}
\label{8}
\Omega&=& J_1\eta_A\eta_B+\frac{J_2}2(\eta_A^2+\eta_B^2)
     +T\ln\left\{\left(\eta_A^2-\frac 12\right)\left(\eta_B^2-
     \frac 12\right)\right\} \nonumber\\
&&-T\ln\left(1+{\rm e}^{-\beta(-2\mu+U+g)}+
  2{\rm e}^{-\beta\left(-\mu+\frac g2\right)}\right)\nonumber\\
&&-T\ln\left(1+{\rm e}^{-\beta(-2\mu+U-g)}+
  2{\rm e}^{-\beta\left(-\mu-\frac g2\right)}\right).
\end{eqnarray}

\subsubsection{$\mu=const$ regime.}

\begin{wrapfigure}[7]{r}{2.6cm}
 \label{fig6}
 \raisebox{-.3cm}[1.cm][.5cm]{\epsfxsize 2.5cm\epsfbox{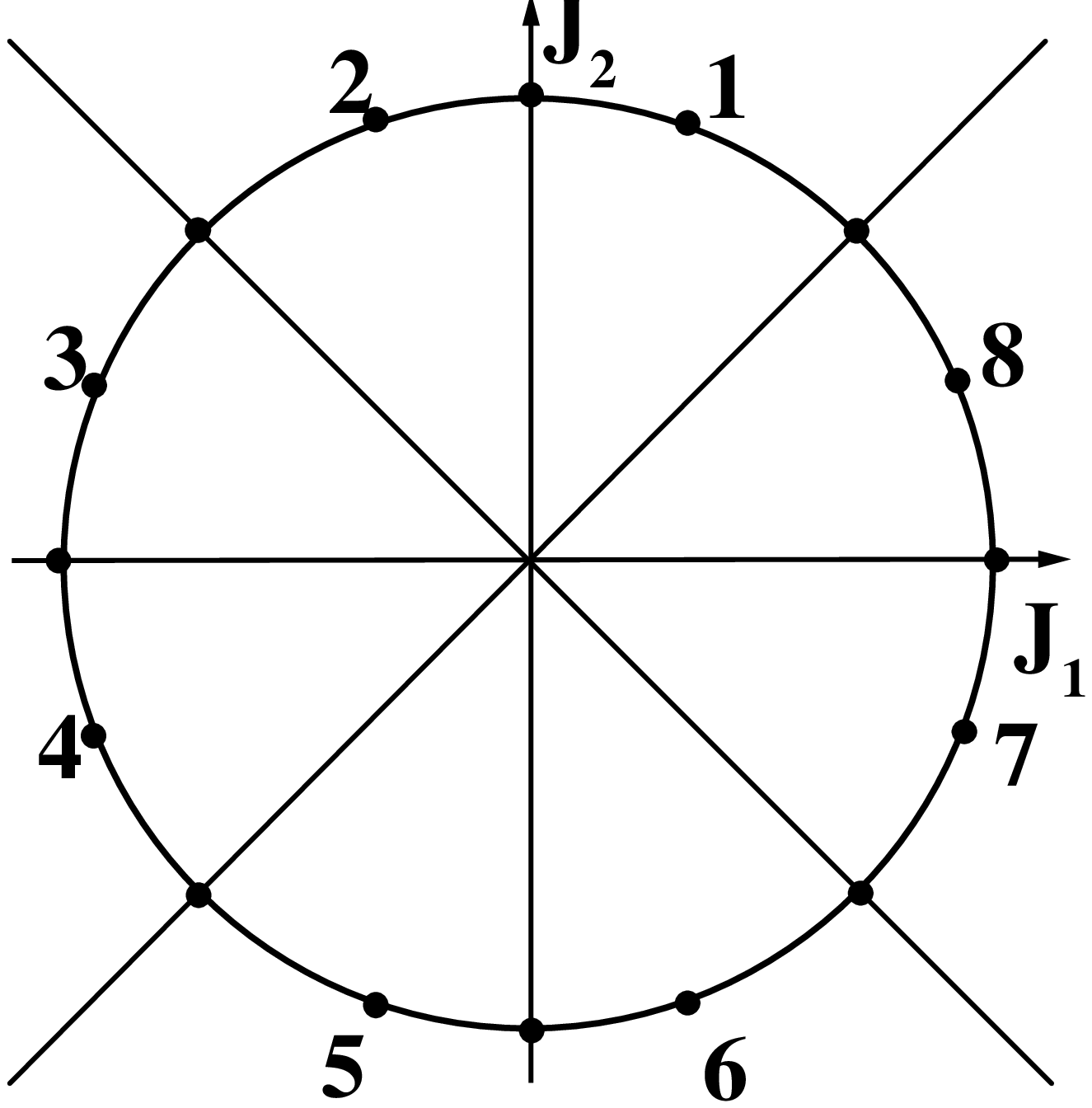}}
\parbox[c]{2.5cm}{Figure 6. $J_1$-$J_2$ area.}
\end{wrapfigure}
 The set of equations for $\eta_{A}$, $\eta_{B}$ is defined by the
mi\-ni\-mum of the grand canonical potential (\ref{8}).
 The expression for the electron mean values is determined from
the thermodynamical relation (\ref{4}).

 The form of the grand canonical potential field dependence $\Omega (h)$
(and therefore the type and number of phase transitions) depends on the
relation between parameters $J_{1}$ and $J_{2}$ values (Fig.~6).
 There are no any specific behaviour when $J_{1}$ and $J_{2}$ are placed in
the regions 5~and~6.
 The case when $J_{1}$ and $J_{2}$ are placed in the domains 1, 7, 8 is
similar to ferroelectric type interaction.

\begin{figure}[htbp]
 \centerline{
  \epsfysize 3.2cm\epsfbox{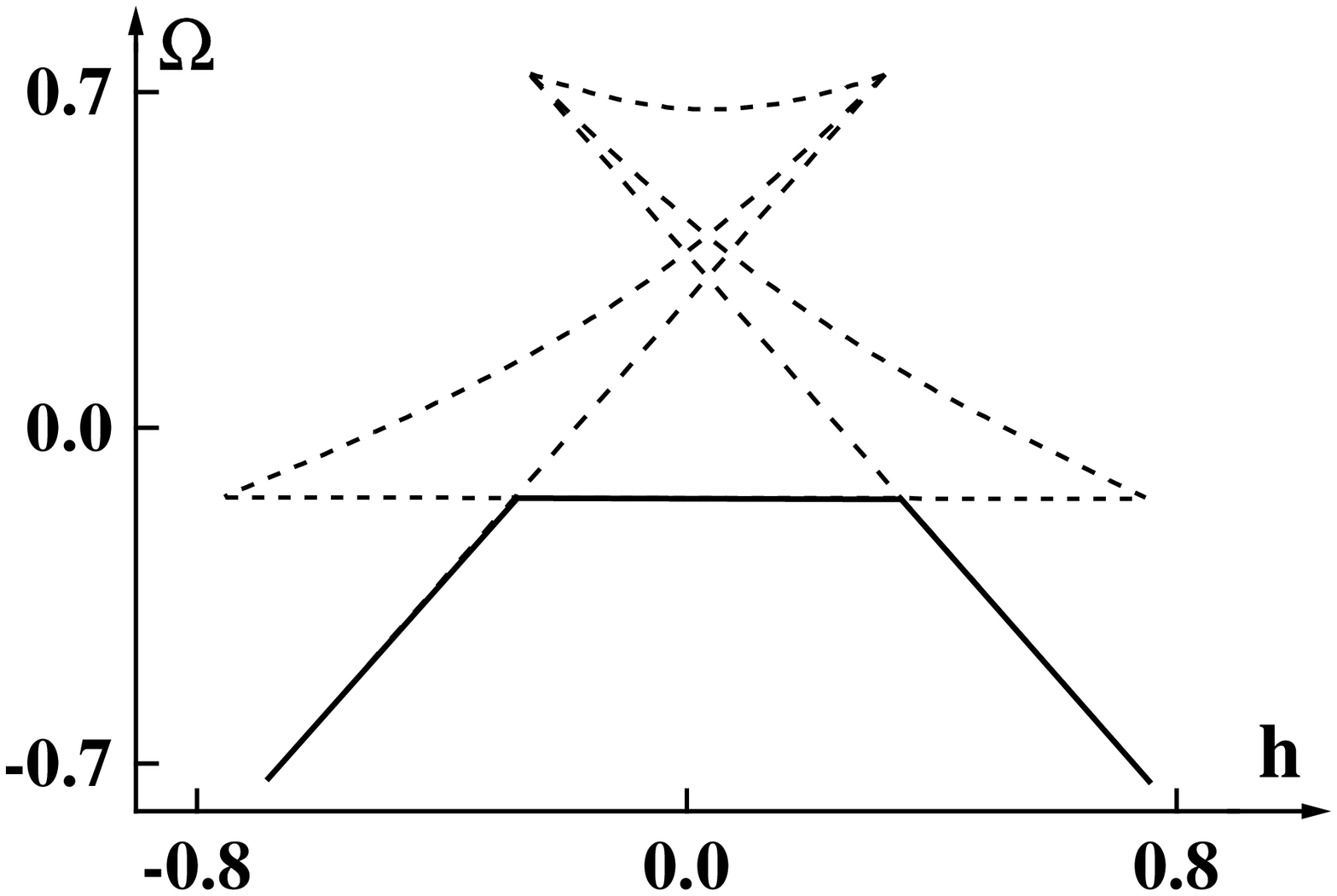}
  \qquad  \epsfysize 3.2cm\epsfbox{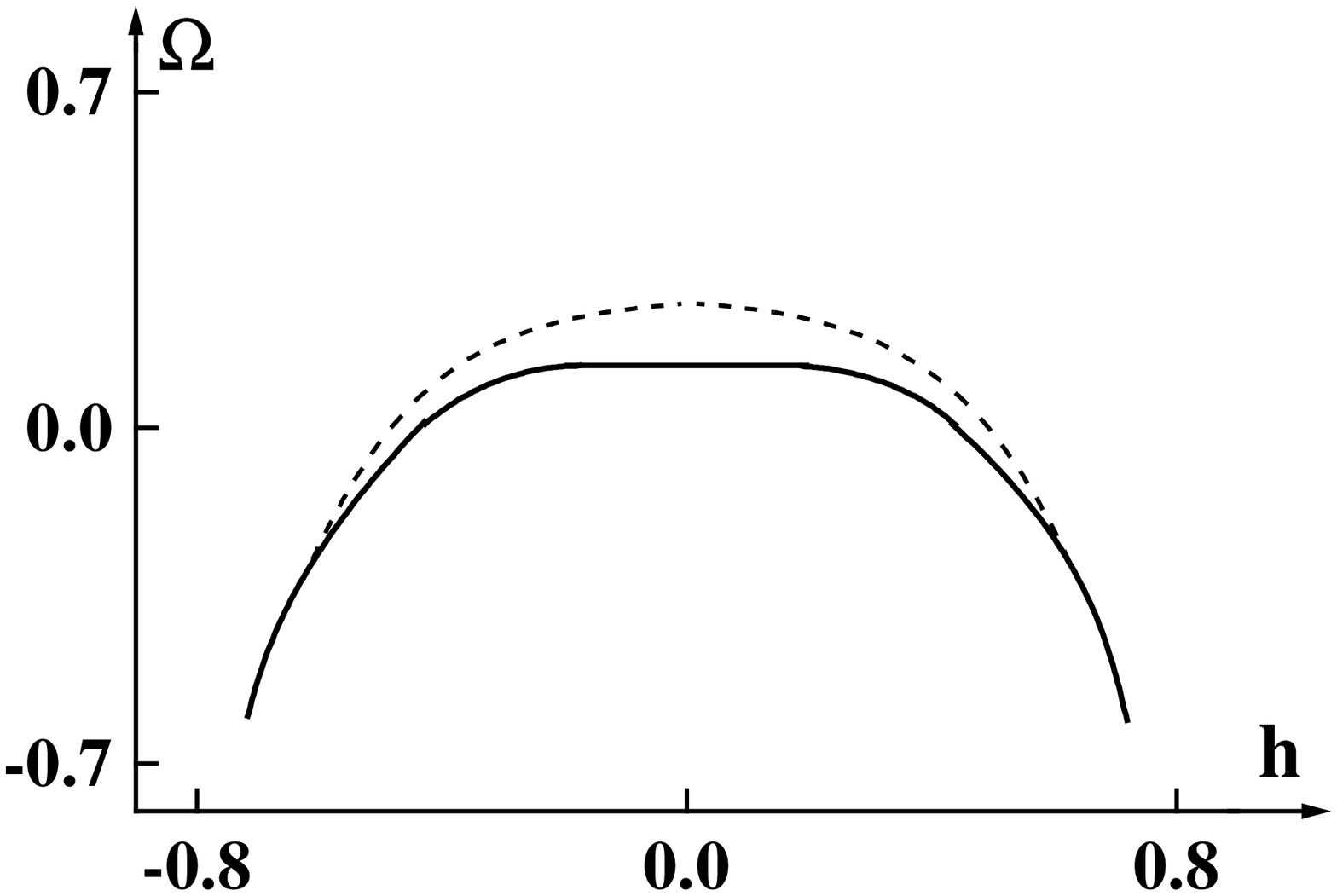}
  }
  Figure 7. Field dependencies of grand canonical potential within the
  2~and~4 regions respectively ($T=0$).
 \label{fig7}
\end{figure}
 The location of $J_{1}$ and $J_{2}$ parameters within the area 4 leads to
the possibility of the two sequential second order phase
transitions: from the ferroelectric phase (FP) with the one
pseudospin mean value to the antiferroelectric phase (AP) and then
to the FP with the another order parameter value (Fig.~7).

\begin{wrapfigure}[10]{r}{4.4cm}
 \label{fig8}
  \raisebox{-1.5cm}[1.cm][1.65cm]{\epsfxsize 4.3cm\epsfbox{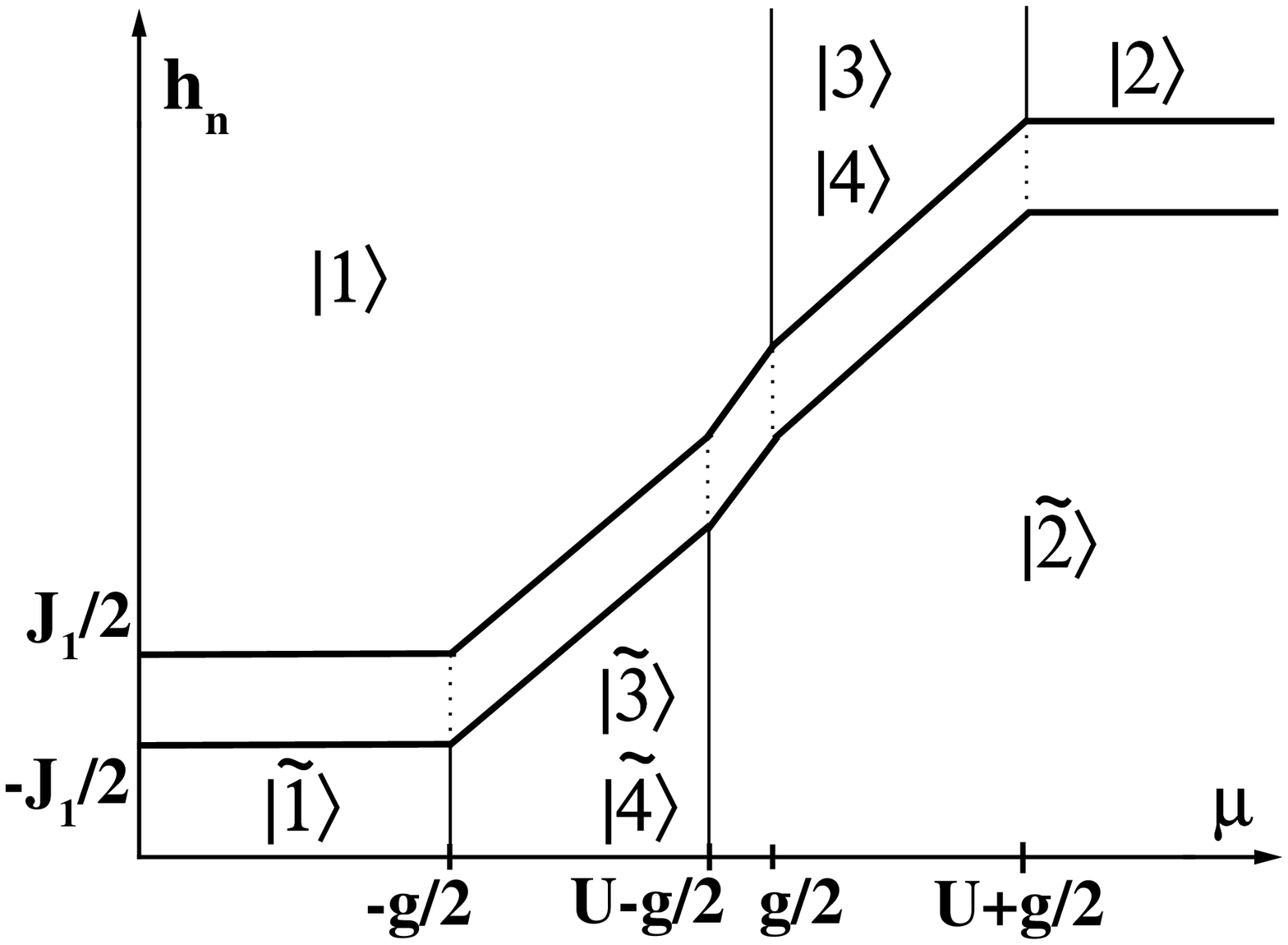}}
\parbox[c]{4.3cm}{Figure 8. $\mu -h$ phase diagram ($T=0$).}
\end{wrapfigure}
 The case when $J_{1}$ and $J_{2}$ belong to the region 2, 3 is
shown in Fig.~7 and the corresponding phase diagram $\mu$--$h$ in
Fig.~8.

 One can see that two first order phase transitions with the change of the
field $h$ and/or chemical potential $\mu$ take place.
 With respect to ferroelectric type interaction between pseudospins
(Fig.~\ref{fig2}) the phase coexistence curve is split and one
obtains the range (the range width is equal $J_1$) where the AP exists
(Fig.~8).
 With the temperature increase the first order phase transitions transform
into the second order phase transitions.
 Hence the possibility of the first order phase transition from FP into AP
and then the second order phase transition from AP into FP exist
with the temperature increase for the narrow range of $h$ values.

\subsubsection{$n=const$ regime.}

 As it was mentioned above, in the case of the fixed value of the electron
concentration (regime $n=$const) the first order phase transition
transform into the phase separation.

 In Fig.~9 one present the phase diagram $n$--$h$ when $J_1$, $J_2$
are placed in the region 2.
 Within the area surrounded by the lines the phase separation into
the regions with different concentrations and phases FP (solid
lines) and AP (dotted lines) takes place.
 Outside of these boundaries (which surround the phase separated states)
the state with the space homogeneity of electron concentration
(FP) is stable.
 Between the boundaries one have the AP.

\begin{figure}[htbp]
 \centerline{
  \epsfysize 3.2cm\epsfbox{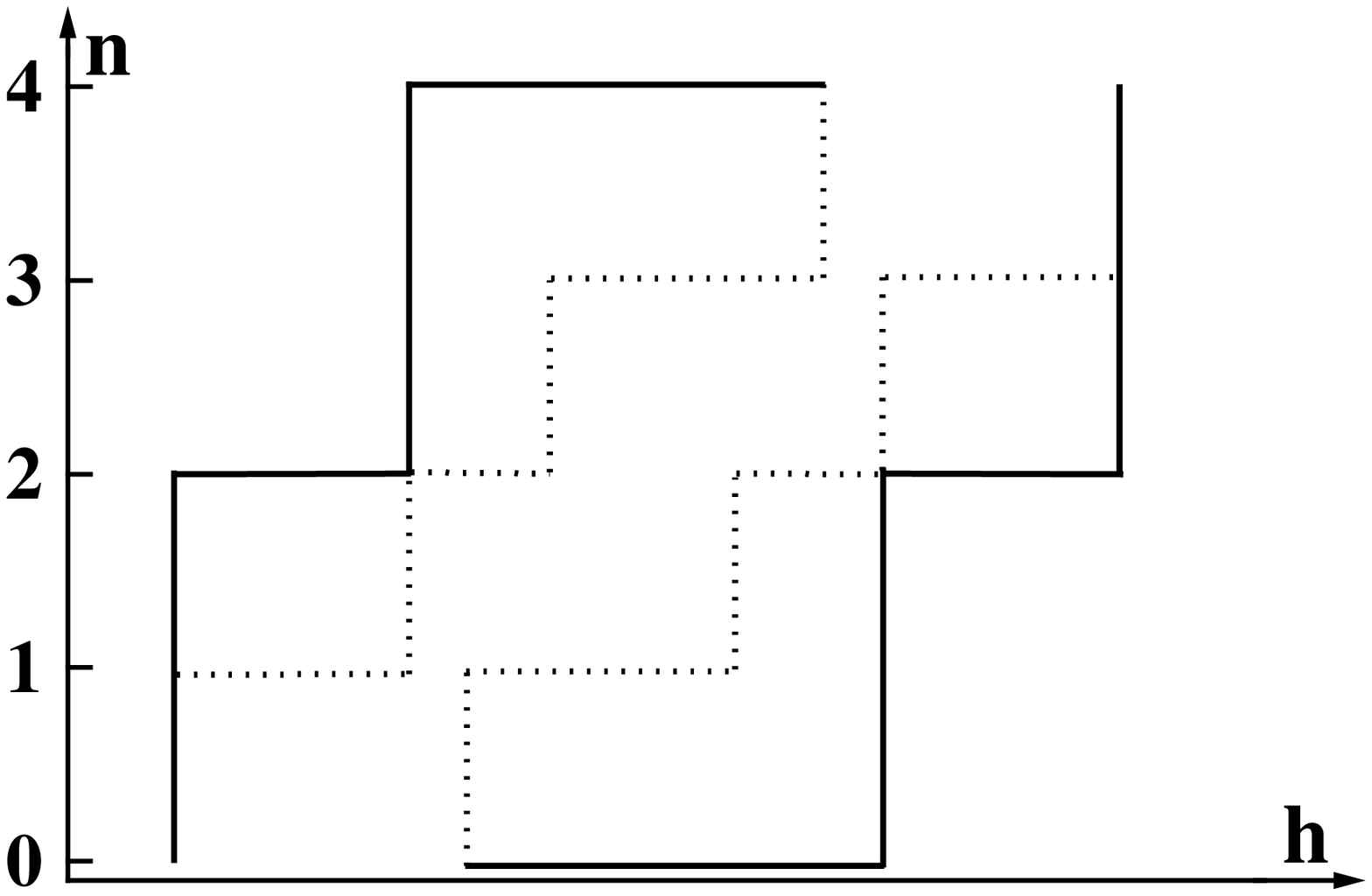}
  \qquad  \epsfysize 3.2cm\epsfbox{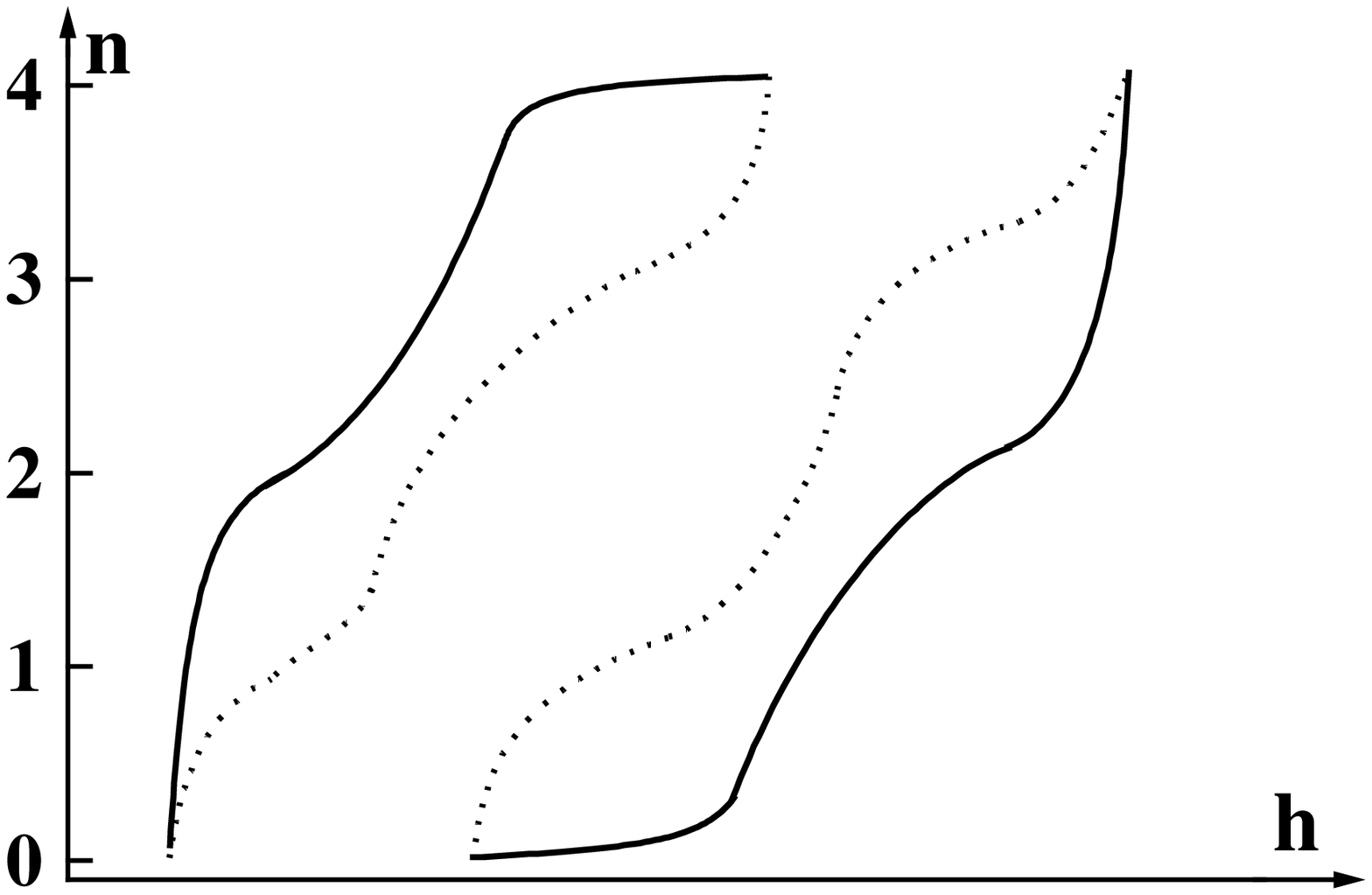}
  }
  \vspace*{.1cm}
  \centerline{
  Figure 9. $n-h$ phase diagram ($T=0$ and $T\neq 0$).
  }
 \label{fig9}
\end{figure}
 We would like to remind that in the $\mu=$const regime with the
temperature increase the first order phase transitions transform
into the second order ($J_1$, $J_2$ $\in$ domain 2,3).
 On the other hand, in the $n=$const regime this correspond to the
narrowing of the range of the phase separated states and transform
into the second order phase transition curves.
 Then the phase transition curves approach one to another and, finally,
disappear at the certain value of temperature.

 The location of $J_1$, $J_2$ within the area 4 leads to the possibility of
the two second order phase transitions with the change of the field
(similar to the $\mu=$const regime).

\section{Interaction between pseudospins via conduct\-ing el\-ec\-tron.}

 In the $U=0$ and $J_{ij}=0$ limit operator (\ref{1})
can be transformed to the Hamiltonian of the electron subsystem of
binary alloy in the case of equilibrium disorder.
 Model (\ref{1}) is close to the
Falicov-Kimball (FK) model but differ in thermodynamic equilibrium
conditions, i.e. in a way how self-consistency is achieved
($\langle S^z\rangle ={\rm const}$ for the FK model and $h={\rm
const}$ for the PE one).

 In the present part of work we propose
(for the case of the $U=0$ and $J_{ij}=0$ limit)
the self-consistent scheme for calculation of mean values of
pseudospin and particle number operators, grand canonical potential
as well as correlation functions.
 The approach is based on the GRPA with the inclusion of the mean field
corrections.
 The possibilities of the phase separation and chess-board phase
appearance are investigated~\cite{7}.

 The calculation is performed in the strong coupling case ($g\gg t$) using
of single-site states as the basic one.
 The formalism of electron creation (annihilation) operators
$
 a_{i\sigma}=b_{i\sigma}P^+_i,
$
$
 \tilde{a}_{i\sigma}=b_{i\sigma}P^-_i
$
($P^{\pm}_i=\frac 12\pm S^z_i$) acting at a site with the certain
pseudospin orientation is introduced.
 Expansion of the calculated quantities in terms of electron transfer
leads to the infinite series of terms containing the averages of
the $T$-products of the $a_{i\sigma}$, $\tilde{a}_{i\sigma}$
operators.
 The evaluation of such averages is made using the corresponding
Wick's theorem.
 The averages of the products of the projection operators
 $P^{\pm}_i$ are expanded in semi-invariants \cite{7}.

 Nonperturbated electron Green function is equal to
\begin{equation}
 \label{9}
 g(\omega_n)=\langle g_i(\omega_n)\rangle;\quad
 g_i(\omega_n)=\frac{P^+_i}{{\rm i}\omega_n-\varepsilon}+
 \frac{P^-_i}{{\rm i}\omega_n-\tilde{\varepsilon}},
\end{equation}
where
$
 \varepsilon =-\mu+g/2 ,\quad
 \tilde{\varepsilon} =-\mu-g/2\quad
$
are single-site energies.
 Single-electron Green function (calculated in Hubbard-I type approximation)
is
$
 \raisebox{-0.13cm}{\epsfysize .4cm\epsfbox{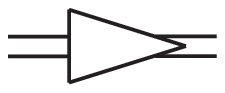}}
$
$
 =G_{\bk}(\omega_n)
$
$
=\big(g^{-1}(\omega_n)-t_{\bk}\big)^{-1}
$
and its poles determine the electron spectrum
\begin{equation}
 \label{10}
 \varepsilon_{\rs {I},\rs {II}}(t_{\bk})=
 \frac{1}{2}(2E_0+t_{\bk})\pm \frac{1}{2}
 \sqrt{g^2+4t_{\bk}\langle S^z \rangle g +t^2_{\bk}}\, .
\end{equation}

 In the adopted approximation the diagrammatic series for the pseudospin mean
value can be presented in the form
\begin{equation}
 \label{11}
 \langle S^z\rangle= \raisebox{-.24cm}{\epsfysize 1.3cm\epsfbox{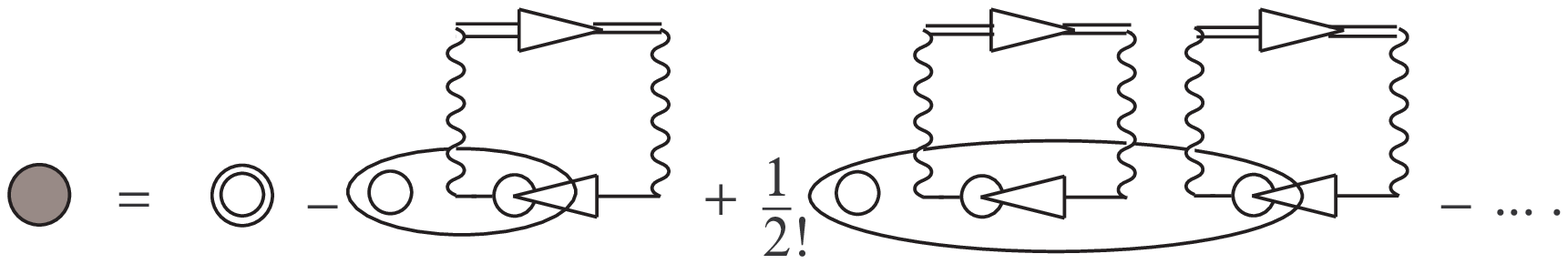}}
\end{equation}
 Here we use the following diagrammatic notations:
$
 \raisebox{-.13cm}{\epsfysize .4cm\epsfbox{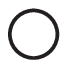}}
 \,-\, S^z,
$
$
 \raisebox{-.13cm}{\epsfysize .4cm\epsfbox{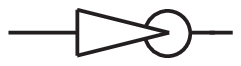}}
$
$
 \,-\, g_i(\omega_n),
$
wavy line is the Fourier transform of hopping $t_{\bk }$.
 Semi-in\-va\-ri\-ants are represented by ovals and contain the
$\delta$-symbols on site indexes.
  In the spirit of the traditional mean field approach \cite{7} the
 renormalization of the basic semi-invariant by the insertion of independent
 loop fragments is taken into account in (\ref{11}).

 The analytical expression for the loop is the following:
\begin{eqnarray}
\label{12}
 \hspace{-2em}
 \raisebox{-.4cm}{\epsfysize 1.1cm\epsfbox{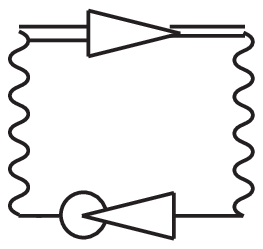}}
 &&=\frac{2}{N}\sum_{n,\bk}t^2_{\bk}
  \frac{\left(P^+_i({\rm i}\omega_n-\varepsilon)^{-1}+
              P^-_i({\rm i}\omega_n-\tilde{\varepsilon})^{-1}
        \right)}
 {g^{-1}(\omega_n)-t_{\bk}}\nonumber\\
 &&=\beta(\alpha_1P^+_i+\alpha_2P^-_i).
\end{eqnarray}

 It should be noted that within the self-consistent scheme
of the GRPA, the chain fragments form the single-electron Green
function in the Hubbard-I approximation and in the sequences of
loop diagrams in the expressions for grand canonical potential
$\Omega$ and pair correlation functions ($\langle
S^z_iS^z_j\rangle$, $\langle S^z_in_j\rangle$, $\langle
n_in_j\rangle$) the connections between any two loops by more than
one semi-invariant are omitted.
 This procedure includes the renormalization of the higher order semi-invariants,
which is similar to the one given by expression (\ref{11}).

From (\ref{11}) and (\ref{12}) follows the equation for pseudospin
mean value
\begin{equation}
 \label{13}
 \langle S^z\rangle=
 \frac{1}{2}\tanh\left\{\frac{\beta}{2}
 (h+\alpha_2(\langle S^z\rangle)-\alpha_1(\langle S^z\rangle))+
 \ln{\frac{1+{\rm e}^{-\beta\varepsilon}}
 {1+{\rm e}^{-\beta\tilde{\varepsilon}}}}
 \right\}.
\end{equation}
  The grand canonical potential in the considered approximation
has the form:
\begin{eqnarray*}
 \label{14}
  &&
  \hspace*{-1.4pc}\!
  \Delta\Omega{=}\Omega{-}\Omega\Big|_{t=0}\!{=}
  {-}\frac{2}{N\beta}\sum_{\bk}
  \ln\frac{(\cosh\frac{\beta}{2}\varepsilon_{\rs I}(t_{\bk}))
  (\cosh\frac{\beta}{2}\varepsilon_{\rs II}(t_{\bk}))}
  {(\cosh\frac{\beta}{2}\varepsilon)
  (\cosh\frac{\beta}{2}\tilde{\varepsilon})}
  {+}\langle S^z\rangle(\alpha_2{-}\alpha_1) \\
 &&\hspace*{-1.6pc}
  {-}\frac{1}{\beta}\ln \cosh\left\{
  \frac{\beta}{2}(h{+}\alpha_2{-}\alpha_1){+}\ln\frac
  {1{+}{\rm e}^{{-}\beta\varepsilon}}
  {1{+}{\rm e}^{{-}\beta\tilde{\varepsilon}}}
  \right\}
  {+}\frac{1}{\beta}\ln \cosh\left\{
  \frac{\beta}{2} h{+}\ln\frac
  {1{+}{\rm e}^{{-}\beta\varepsilon}}
  {1{+}{\rm e}^{{-}\beta\tilde{\varepsilon}}}
  \right\}
  \nonumber
\end{eqnarray*}
 The solution of equation (\ref{13}) and calculation of potential $\Omega$
were performed numerically.

 In the $\mu={\rm const}$ regime (in the uniform case)
there exists a possibility of the first order phase
transition with the jump of the pseudospin mean value and
reconstruction of the electron spectrum \cite{7}.

 In the $n={\rm const}$ regime one can see
(Fig.~10) that the regions with ${\rm d}\mu /{\rm d}n\leq 0$,
where states with a homogeneous distribution of particles are
unstable, exist.
 This corresponds to the phase separation into the states
with different electron concentrations and pseudospin mean values.
 In the phase separated region the free energy as a function of $n$
deflects up (Fig.~10) and concentrations of the separated phases
are determined by the tangent line touch points (these points are
also the points of binodal lines which are determined according to
the Maxwell rule from the function $\mu (n)$).
\begin{figure}[htbp]
 \centerline{
  {\epsfysize 3.6cm\epsfbox{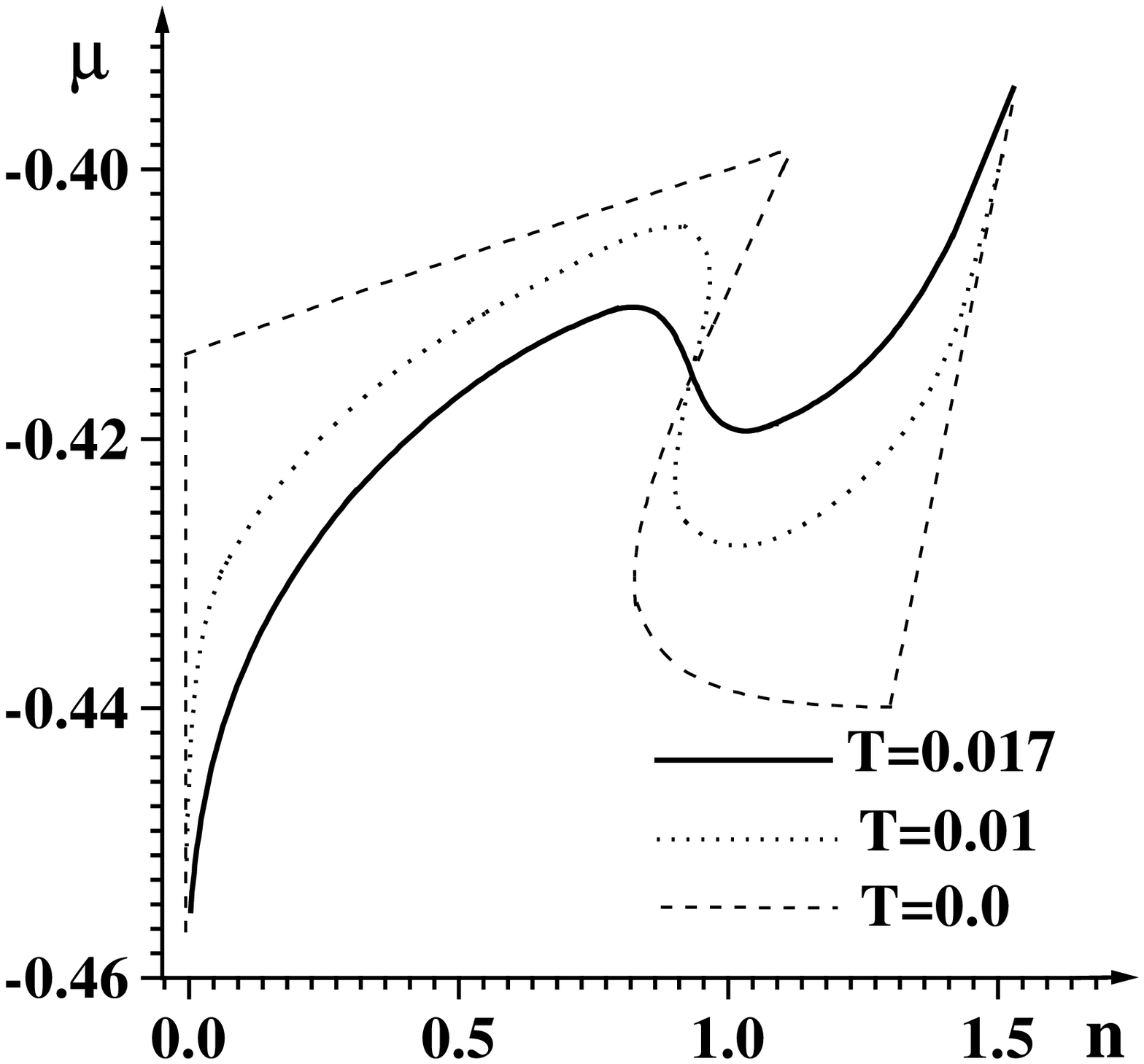}}
  \qquad  {\epsfysize 3.6cm\epsfbox{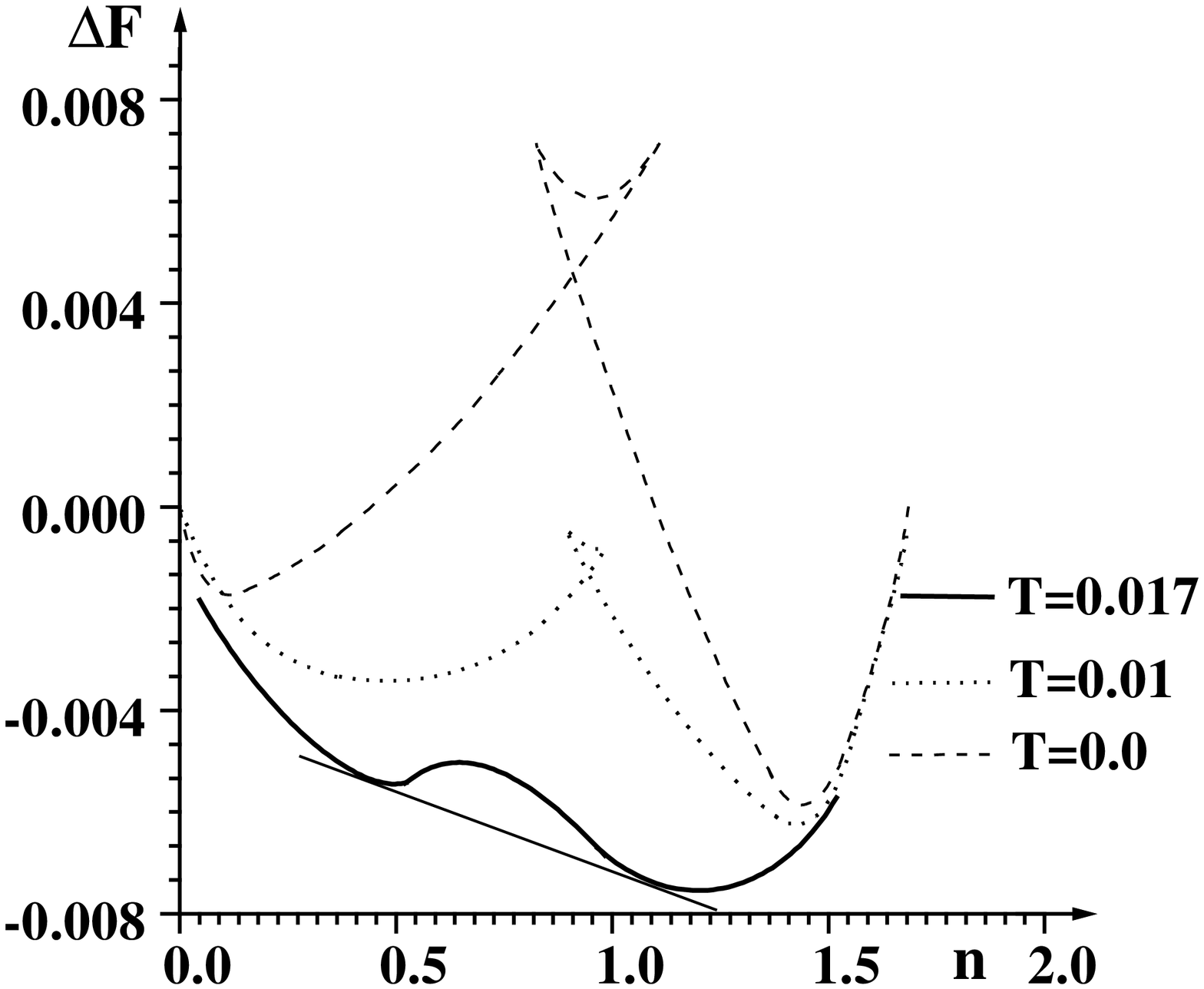}}
  }
  \vspace*{.2cm}
  Figure 10. Dependence of the chemical potential $\mu $ on the electron
  concentration $n$ and deviation of the free energy from linear dependence
  $\Delta F=F(n)-[\frac n2 F(2)+(1-\frac n2)F(0)]$
  for different $T$ values. 
 \label{fig10}
\end{figure}

 The analysis of the $\langle S^zS^z\rangle_{\bq}$ correlator temperature behaviour
shows that the high temperature phase become unstable with respect
to fluctuations with ${\bq}\not =0$ for certain values of model
parameters.
 The maximal temperature of instability is achieved for
${\bq}=(\pi,\pi)$ (in the case of square lattice with
nearest-neighbor hopping) and indicates the possibility of phase
transition into a modulated (chess-board) phase.

 The analytical consideration of the chess-board phase within the
framework of the presented above approximation can be performed in
a similar way.

\begin{wrapfigure}[13]{r}{4.5cm}
 \raisebox{-.5cm}[2.7cm][.7cm]{\epsfxsize 4.3cm\epsfbox{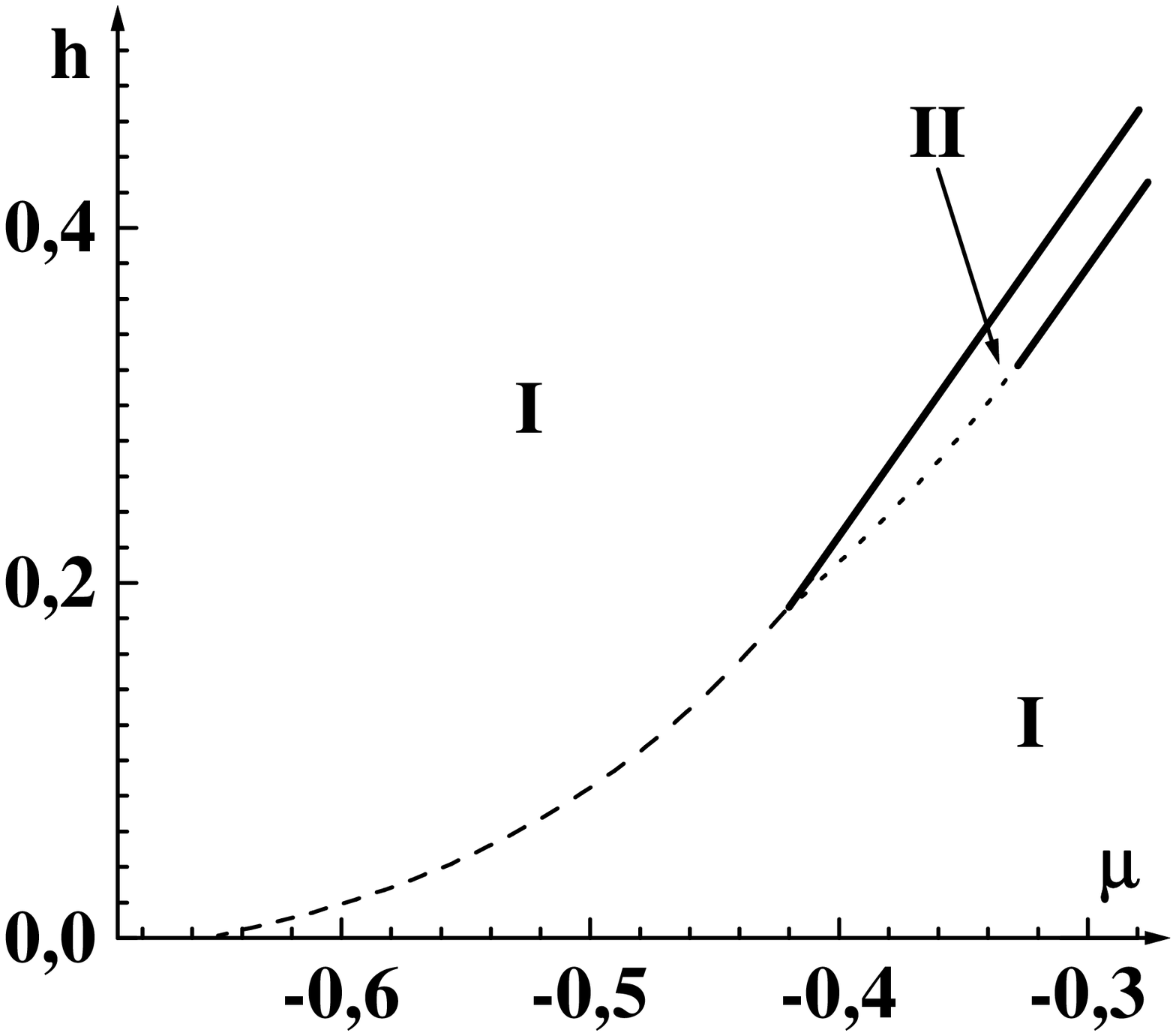}}
 \parbox[c]{4.3cm}{Figure 11. $\mu$-$h$ phase diag\-ram.
                           I~-- uniform phase, II~-- chess-board phase.}
\label{fig11}
\end{wrapfigure}
 From the comparison of the grand canonical potential
$\Omega$ values for uniform and chess-board phases, the
($\mu ,h$) phase diagram is obtained (Fig.~11).
 One can see that chess-board phase exists as an intermediate one
between the uniform phases with the different $\langle S^z\rangle$ and
$n$ values. The transition between different uniform phases
(bistability) is of the first order (Fig.~11, dashed line), while
the transition between the uniform and modulated ones is of the
first (dotted line) or second (solid line) order.

 Appearance of various phases in the con\-si\-de\-red model remind the
situation known for the FK model with a rich phase diagram
\cite{Freericks}.
 However, contrary to this model, an existence of the phase
transitions between uniform phases is possible in our case.
 This results from the another regime of thermodynamic averaging
(fixation of $h$ field which is an analogous to the chemical
potential in the FK model).

$$ $$
 Tabunshchyk K.V. e-mail: tkir@icmp.lviv.ua
\end{lecture}
\end{document}